\documentclass[12pt]{cernart}

\usepackage{epsfig}
\usepackage{graphicx} 
\usepackage{amsmath,amssymb}
\include{defin}
\usepackage{amssymb}

\begin{document}
\begin{titlepage}
\docnum{CERN-PH-EP/2006-032}
\date{3~October~2006}
\title{\bf \Large Measurement of the branching ratios of the decays $\Xi^{0}\rightarrow \Sigma^{+} e^{-} \overline{\nu}_{e}$ 
and $\overline{\Xi^{0}}\rightarrow \overline{\Sigma^{+}} e^{+} \nu_{e}$}

\begin{Authlist}
\begin{center}
{\bf NA48/1 Collaboration}
\  \\[0.2cm] 
 %
%
J.R.~Batley,
G.E.~Kalmus\footnotemark[1],
C.~Lazzeroni,
D.J.~Munday,
M.~Patel,
M.W.~Slater,
S.A.~Wotton \\
{\em \small Cavendish Laboratory, University of Cambridge, Cambridge, CB3 0HE,
UK\footnotemark[2]} \\[0.2cm] 
 R.~Arcidiacono,
 G.~Bocquet,
 A.~Ceccucci,
 D.~Cundy\footnotemark[3],
 N.~Doble\footnotemark[4],
 V.~Falaleev,
 L.~Gatignon,
 A.~Gonidec,
 P.~Grafstr\"om,
 W.~Kubischta,
F.~Marchetto\footnotemark[5],
 I.~Mikulec\footnotemark[6],
 A.~Norton,
 B.~Panzer-Steindel,
P.~Rubin\footnotemark[7],
 H.~Wahl\footnotemark[8] \\
{\em \small CERN, CH-1211 Gen\`eve 23, Switzerland} \\[0.2cm] 
E.~Goudzovski,
P.~Hristov\footnotemark[9],
V.~Kekelidze,
L.~Litov,
D.~Madigozhin,
N.~Molokanova,
Yu.~Potrebenikov,
S.~Stoynev,
A.~Zinchenko\\
{\em \small Joint Institute for Nuclear Research, Dubna, Russian    Federation} \\[0.2cm] 
E.~Monnier\footnotemark[10],
E.C.~Swallow\footnotemark[11],
R.~Winston\footnotemark[12]\\
{\em \small The Enrico Fermi Institute, The University of Chicago, Chicago, Illinois, 60126, U.S.A.}\\[0.2cm]
 R.~Sacco\footnotemark[13],
 A.~Walker \\
{\em \small Department of Physics and Astronomy, University of    Edinburgh, JCMB King's Buildings, Mayfield Road, Edinburgh,    EH9 3JZ, U.K.} \\[0.2cm] 
%
W.~Baldini,
P.~Dalpiaz,
P.L.~Frabetti,
A.~Gianoli,
M.~Martini,
F.~Petrucci,
M.~Scarpa,
M.~Savri\'e \\
{\em \small Dipartimento di Fisica dell'Universit\`a e Sezione dell'INFN di Ferrara, I-44100 Ferrara, Italy} \\[0.2cm] 
%
%
A.~Bizzeti\footnotemark[14],
M.~Calvetti,
G.~Collazuol\footnotemark[15],
E.~Iacopini,
M.~Lenti,
G.~Ruggiero\footnotemark[9],
M.~Veltri\footnotemark[16] \\
{\em \small Dipartimento di Fisica dell'Universit\`a e Sezione dell'INFN di Firenze, I-50125 Firenze, Italy} \\[0.2cm] 
%
%
M.~Behler,
K.~Eppard,
 M.~Eppard\footnotemark[9],
 A.~Hirstius\footnotemark[9],
 K.~Kleinknecht,
 U.~Koch,
L.~Masetti, 
P.~Marouelli,
U.~Moosbrugger,
C.~Morales Morales,
 A.~Peters\footnotemark[9],
 R.~Wanke,
 A.~Winhart \\
{\em \small Institut f\"ur Physik, Universit\"at Mainz, D-55099 Mainz,
Germany\footnotemark[17]} \\[0.2cm] 
A.~Dabrowski,
T.~Fonseca Martin\footnotemark[9],
M.~Velasco \\
{\em \small Department of Physics and Astronomy, Northwestern University, Evanston Illinois 60208-3112, U.S.A.}
 \\[0.2cm] 
G.~Anzivino,
P.~Cenci,
E.~Imbergamo,
G.~Lamanna\footnotemark[18],
P.~Lubrano,
A.~Michetti,
A.~Nappi,
M.~Pepe,
M.C.~Petrucci,
M.~Piccini\footnotemark[9],
M.~Valdata \\
{\em \small Dipartimento di Fisica dell'Universit\`a e Sezione    dell'INFN di Perugia, I-06100 Perugia, Italy} \\[0.2cm] 
%
%
 C.~Cerri,
F.~Costantini,
 R.~Fantechi,
 L.~Fiorini\footnotemark[19],
 S.~Giudici,
 I.~Mannelli,
G.~Pierazzini,
 M.~Sozzi \\
{\em \small Dipartimento di Fisica, Scuola Normale Superiore e Sezione dell'INFN di Pisa, I-56100 Pisa, Italy} \\[0.2cm] 
%
%
C.~Cheshkov,
J.B.~Cheze,
 M.~De Beer,
 P.~Debu,
G.~Gouge,
G.~Marel,
E.~Mazzucato,
 B.~Peyaud,
 B.~Vallage \\
{\em \small DSM/DAPNIA - CEA Saclay, F-91191 Gif-sur-Yvette, France} \\[0.2cm] 
M.~Holder,
 A.~Maier,
 M.~Ziolkowski \\
{\em \small Fachbereich Physik, Universit\"at Siegen, D-57068 Siegen,
Germany\footnotemark[20]} \\[0.2cm] 
C.~Biino,
N.~Cartiglia,
M.~Clemencic, 
S.~Goy Lopez,
E.~Menichetti,
N.~Pastrone \\
{\em \small Dipartimento di Fisica Sperimentale dell'Universit\`a e    Sezione dell'INFN di Torino,  I-10125 Torino, Italy} \\[0.2cm] 
 W.~Wislicki,
\\
{\em \small Soltan Institute for Nuclear Studies, Laboratory for High    Energy
Physics,  PL-00-681 Warsaw, Poland\footnotemark[21]} \\[0.2cm] 
H.~Dibon,
M.~Jeitler,
M.~Markytan,
G.~Neuhofer,
L.~Widhalm \\
{\em \small \"Osterreichische Akademie der Wissenschaften, Institut  f\"ur
Hochenergiephysik,  A-10560 Wien, Austria\footnotemark[22]} \\[1cm] 
\vspace{0.5cm}
\it{Submitted for publication in Physics Letters B.}
\end{center}
\setcounter{footnote}{0}
\footnotetext[1]{Present address: Rutherford Appleton Laboratory,
Chilton, Didcot, OX11 0QX, UK} 
\footnotetext[2]{ Funded by the U.K.    Particle Physics and Astronomy Research Council}
\footnotetext[3]{Present address: Instituto di Cosmogeofisica del CNR di Torino, I-10133 Torino, Italy}
\footnotetext[4]{Also at Dipartimento di Fisica dell'Universit\`a e Sezione dell'INFN di Pisa, I-56100 Pisa, Italy}
\footnotetext[5]{On leave from Sezione dell'INFN di Torino,  I-10125 Torino, Italy}
\footnotetext[6]{ On leave from \"Osterreichische Akademie der Wissenschaften, Institut  f\"ur Hochenergiephysik,  A-1050 Wien, Austria}
\footnotetext[7]{On leave from University of Richmond, Richmond, VA, 23173, 
USA; supported in part by the US NSF under award \#0140230. Present address:
Department of Physics and Astronomy George Mason University, Fairfax, VA 22030A,
USA}
\footnotetext[8]{Also at Dipartimento di Fisica dell'Universit\`a e Sezione dell'INFN di Ferrara, I-44100 Ferrara, Italy}
\footnotetext[9]{Present address: CERN, CH-1211 Gen\`eve 23, Switzerland}
\footnotetext[10]{Present address: Centre de Physique des Particules de Marseille, IN2P3-CNRS, Universit\'e 
de la M\'editerran\'ee, Marseille, France}
\footnotetext[11]{Present address: Department of Physics, Elmhurst College, 
Elmhurst, IL, 60126, USA}
\footnotetext[12]{Also at University of California, Merced, USA}
\footnotetext[13]{Present address: Department of Physics Queen Mary, University
of London, Mile End Road, London E1 4NS, United Kingdom}
\footnotetext[14]{ Dipartimento di Fisica dell'Universit\`a di Modena e Reggio Emilia, via G. Campi 213/A I-41100, Modena, Italy}
\footnotetext[15]{Present address: Scuola Normale Superiore e Sezione 
dell'INFN di Pisa, I-56100 Pisa, Italy} 
\footnotetext[16]{ Istituto di Fisica, Universit\`a di Urbino, I-61029  Urbino, Italy}
\footnotetext[17]{ Funded by the German Federal Minister for    Research and Technology (BMBF) under contract 7MZ18P(4)-TP2}
\footnotetext[18]{Present address: Dipartimento di Fisica 
 e Sezione dell'INFN di Pisa, 
I-56100 Pisa, Italy} 
\footnotetext[19]{Present address: Cavendish Laboratory, University of 
Cambridge, Cambridge, CB3 0HE, U.K.} 
\footnotetext[20]{ Funded by the German Federal Minister for Research and Technology (BMBF) under contract 056SI74}
\footnotetext[21]{Supported by the Committee for Scientific Research grants
5P03B10120, SPUB-M/CERN/P03/DZ210/2000 and SPB/CERN/P03/DZ146/2002}
\footnotetext[22]{Funded by the Austrian Ministry for Traffic and 
Research under the 
contract GZ 616.360/2-IV GZ 616.363/2-VIII, 
and by the Fonds f\"ur   Wissenschaft und Forschung FWF Nr.~P08929-PHY}

\end{Authlist}
\end{titlepage}


\begin{abstract}

From 56 days of data taking in 2002, the NA48/1 experiment observed  
6316 $\Xi^{0}\rightarrow \Sigma^{+} e^{-} \overline{\nu}_{e}$ candidates 
(with the subsequent $\Sigma^{+} \rightarrow p \pi^{0}$ decay) and 555 
$\overline{\Xi^{0}}\rightarrow \overline{\Sigma^{+}} e^{+} \nu_{e}$ candidates
with background contamination of $215 \pm 44$ and $136 \pm 8$ events, 
respectively. From these samples, the branching 
ratios BR($\Xi^{0}\rightarrow \Sigma^{+} e^{-} \overline{\nu}_{e})=
(2.51 \pm0.03_{\mathrm{stat}}\pm 0.09_{\mathrm{syst}})\times10^{-4}$ 
and BR($\overline{\Xi^{0}}\rightarrow \overline{\Sigma^{+}} e^{+} \nu_{e})=
(2.55 \pm 0.14_{\mathrm{stat}}\pm{0.10}_{\mathrm{syst}}) \times 10^{-4}$ 
were measured allowing the determination of the CKM matrix element
$|V_{\mathrm{us}}| = 0.209^{+0.023}_{-0.028}$. Using the Particle Data Group
average for $|V_{\mathrm{us}}|$ obtained in semileptonic kaon decays, we 
measured the ratio $g_1/f_1 = 1.20 \pm 0.05$ of the axial-vector to vector 
form factors.

\end{abstract}




\vspace{0.2cm}

\section{Introduction}
The study of hadron $\beta$-decays gives important information on the 
interplay between the weak interaction and the hadronic structure
determined by the strong interaction.
This information is richer for baryon than for meson semileptonic decays
owing to the presence of three valence quarks as opposed to a
quark-antiquark pair.
In this context, $\Xi^0$ $\beta$-decay represents an extraordinary 
opportunity to test, by analogy with neutron $\beta$-decay, SU(3) 
symmetry and, through the determination of $V_{\mathrm{us}}$, the quark 
mixing model~\cite{cabibbo}. 

In the exact SU(3) symmetry approximation, the ratio
between the axial-vector form factor $g_{1}$ and the vector 
form factor $f_{1}$ for $\Xi^{0}$ $\beta$-decay is equal to the 
one for the decay $n \rightarrow p e^{-} \overline{\nu}_{e}$. Theoretical 
models that incorporate SU(3) symmetry breaking effects give
predictions which, however, differ significantly from each other
\cite{donoghue,carson,krause,anderson,schlumpf,mendieta,mendieta2,ratcliffe}.
Precise tests of SU(3) symmetry breaking effects calculations in semileptonic 
hyperon decays are therefore important in connection with the  
determination of $V_{\mathrm{us}}$, independently from kaon decays.
  
Recently, the
KTeV experiment has obtained the first determination of the $g_{1}/f_{1}$
ratio in $\Xi^{0} \rightarrow \Sigma^{+} e^{-} \overline{\nu}_{e}$ 
decays from the study of the $\Sigma^{+}$ polarization with the 
decay $\Sigma^{+} \rightarrow p\pi^{0}$ and the $e^{-}-\overline{\nu}_{e}$ 
correlation~\cite{ktevff}.
Their result, based on the observation of 487 events, 
is consistent 
with exact SU(3) symmetry: $g_{1}/f_{1} = 1.32^{+0.21}_{-0.17 \mathrm{stat}} \pm 0.05_{\mathrm{syst}}$. Previously, the same Collaboration 
published the value of the 
branching ratio 
BR($\Xi^{0}\rightarrow \Sigma^{+} e^{-} \overline{\nu}_{e})=
(2.71 \pm0.22_{\mathrm{stat}}\pm 0.31_{\mathrm{syst}})\times10^{-4}$ 
from a sample of 176 events after background subtraction~\cite{ktev2}.   

In the present work, the $\Xi^{0}$ and 
$\overline{\Xi^{0}}$ $\beta$-decay modes have been investigated with 
significantly improved statistics as compared to
previous experiments. The corresponding branching ratios were determined 
relative to the decay
channels $\Xi^0 \rightarrow \Lambda \pi^0$ and 
$\overline{\Xi^{0}} \rightarrow \overline{\Lambda} \pi^0$, respectively, 
allowing the measurement of the matrix element $|V_{\mathrm{us}}|$.
Conversely, using as input parameter the current experimental value 
for $V_{\mathrm{us}}$ from semileptonic kaon decays, 
the form factor $g_{1}/f_{1}$ was determined.

\section{Beam and detector}

The main goal of the NA48/1 experiment is 
the study of very rare $K_{S}$ decay modes and neutral hyperon decays.
A detailed description of the beam line 
and the detector 
can be found in~\cite{epsi}. Only 
the aspects relevant to this measurement are reviewed here.

\subsection{Beam}
The experiment was performed at the CERN SPS accelerator and used a
400~GeV/c proton beam impinging on a Be target 
to produce a neutral beam. The spill length was 4.8~s out of a 16.2~s 
cycle time. The proton intensity was fairly constant during the spill with a 
mean of $5 \times 10^{10}$  particles per pulse.

%

For this measurement, only
the $K_{S}$ target station of the NA48 double 
$K_{S}/K_{L}$ beam line~\cite{epsi} 
was used to produce the neutral beam. In this configuration,
the $K_L$ beam was blocked and an additional sweeping magnet 
was installed to deflect charged particles away from 
the defining section of the $K_S$ collimators.
To reduce the number of photons in the neutral beam originating
primarily from
$\pi^0$ decays, a 24~mm thick platinum absorber was placed in the 
beam between the target and the collimator.
A pair of coaxial collimators, having a total thickness of 5.1~m, the axis 
of which formed an angle of 4.2~mrad
to the proton beam direction, selected a beam of 
neutral long-lived particles ($K_S$, $K_L$,
$\Lambda^0$, $\Xi^0$, $n$ and $\gamma$). 
The aperture of the defining collimator, 5.03~m 
downstream of the target, was a circle with 1.8~mm radius.
The target position and the production angle where chosen in such a way
that the beam axis was hitting the center of the electromagnetic calorimeter.  
  
In order to minimize the 
interaction of the neutral beam with air,
the collimator was immediately followed by a $90$~m long 
evacuated tank terminated by a 0.3$\%$ $X_0$ thick Kevlar 
window. The NA48 detector was located downstream of this region 
in order to collect the products of the particles decaying 
in the volume contained by the tank.

On average,  about $1.4 \times 10^{4}$  $\Xi^{0}$ per spill, with an energy 
between 70 and 220~GeV, decayed in
the fiducial decay volume.

\subsection{Tracking}
\label{tracking}
The detector included a spectrometer 
housed in a helium gas volume
with two drift chambers before (DCH1, DCH2) and two after (DCH3, DCH4)
a dipole magnet with a horizontal transverse momentum kick of 265~MeV/c. 
Each chamber had four views ({\it x, y, u, v}), each of which had two 
sense wire planes. In DCH1, DCH2 and DCH4, all wire planes were
instrumented while in the drift chamber located just downstream 
of the magnet (DCH3), only vertical and horizontal wire planes were read out. 
The resulting space points were reconstructed with a
resolution of about 150~$\mu$m in each projection.
The spectrometer momentum resolution could be parameterized as:
\begin{equation}
\sigma_p /p = 0.48 \% \oplus 0.015\% \times p 
\end{equation}

\noindent where $p$ is in GeV/c. This
resulted in a resolution of about 1~MeV/c$^2$ 
when reconstructing the $\Lambda$ mass in $\Lambda \rightarrow p \pi^-$
decays.
The track time resolution was about 1.4~ns.
%

%
   
\subsection{Calorimetry}

The electromagnetic showers were detected and measured with a 27 
radiation-length deep liquid krypton calorimeter (LKr)   
read out in longitudinal cells with a $\sim2\times2$~cm$^2$ cross-section.

The energy resolution 
was given by~\cite{unal}:

\begin{equation}
\sigma(E) / E = \frac{3.2\%}{\sqrt{E}} \oplus \frac{9 \% }{E} \oplus 0.42 \%
\end{equation}
where $E$ is in GeV. The transverse position resolution for a single photon
of energy larger than 20~GeV was better than 1.3 mm and
the corresponding mass resolution at the $\pi^0$ mass was
about 1~MeV/c$^2$.
The time resolution of the calorimeter for a single shower was
better than 300~ps.

A scintillating fiber hodoscope (NHOD), placed inside the LKr calorimeter
at a depth of about 9.5~$X_0$ near the shower maximum, was used for trigger
efficiency measurements.    

The LKr calorimeter was followed by a hadron calorimeter (HAC) consisting 
of an iron-scintillator sandwich, 6.7 nuclear interaction lengths thick. The 
HAC provided a raw
measurement of the energy for hadron showers and it was only used at the 
first trigger level.

\subsection{Scintillator Detectors}
A scintillator hodoscope (CHOD) was located between the
spectrometer and the calorimeter. It consisted of two planes, segmented
in horizontal and vertical strips and arranged in four quadrants.
The CHOD time resolution was better than 200~ps for 2-track events.  
Muon counters made of three planes of scintillator, each shielded by 
an iron wall, were placed at the downstream end of the apparatus.
Seven rings of scintillation counters (AKL), placed around the evacuated decay 
volume and around the helium tank of the charged particle spectrometer, were
used to veto activity outside the acceptance region of the detector 
determined by the LKr calorimeter. 

\section{Trigger}
The trigger system used for the on-line selection of $\Xi^{0}$ $\beta$-decays
consisted of three levels of logic. Level 1 (L1) was based on logic 
combinations of fast signals coming from various sub-detectors. It required  
hits in the CHOD and in the first drift chamber compatible with at least one 
and two tracks respectively, no hit in the AKL veto system and a minimum 
energy deposition in the calorimeters. This last requirement was 
15~GeV for the energy reconstructed in the LKr calorimeter or 30~GeV
for the summed energy in the electromagnetic and hadronic calorimeters.
The output rate of the L1 stage was about 50\,kHz. The average L1 
efficiency, measured
with $\Xi^0 \rightarrow \Lambda \pi^0$ events of energy greater than 
70\,GeV, was found to be $98.65 \pm 0.03\%$.

\begin{figure}[htbp]
  \centering
    \includegraphics[width=10cm]{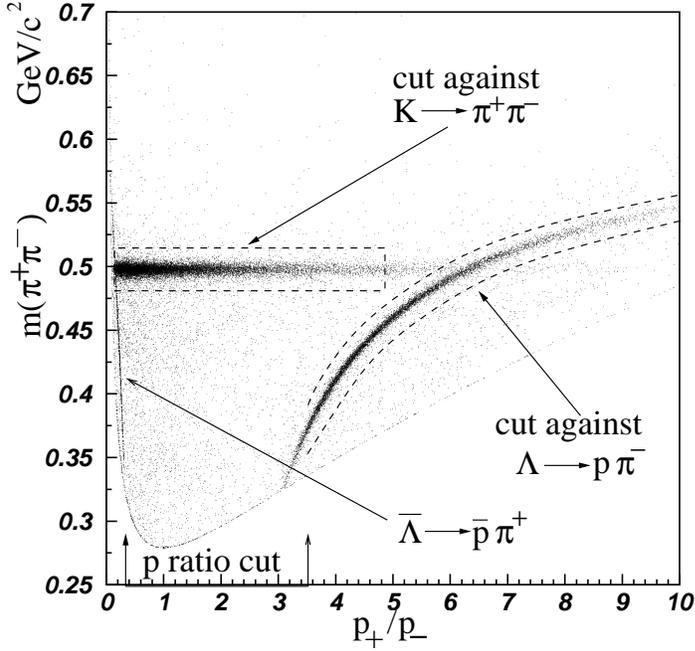}
  \caption{Reconstructed $\pi^{+}\pi^{-}$ invariant mass versus momentum ratio
of positive to negative particles for 2-track vertices as selected by the
L1 trigger. Events coming from $K_{S} \rightarrow \pi^{+}\pi^{-}$ and 
$\Lambda \rightarrow p \pi^{-}$ decays are clearly visible. 
$\overline{\Lambda}\rightarrow \overline{p} \pi^{+}$ decays are located in 
the $p_{+}/p_{-}<0.3$ region. The kinematical 
regions rejected by the L2 trigger are also shown.  
}
  \label{kincut}
\end{figure}

Level 2 (L2) consisted of 300\,MHz processors that reconstructed
 tracks and vertices from hits in the drift chambers 
and computed relevant physical quantities. The L2 trigger required
at least two tracks with a closest distance of approach of less than 8~cm 
in space and a transverse separation greater than 5~cm in the 
first drift chamber. Since the signature of the $\Xi^0$ $\beta$-decay
involves the detection of an energetic proton from the 
subsequent $\Sigma^+ \rightarrow p\pi^0$ decay, the ratio between the higher 
and the lower of the two track momenta was required to be larger 
than 3.5. Rejection of the overwhelming $\Lambda \rightarrow p\pi^-$ and 
$K_S \rightarrow \pi^{+}\pi^{-}$ decays was 
achieved by applying stringent invariant mass cuts
according to the corresponding event hypotheses, $p\pi$ or 
$\pi\pi$ (see Fig.~\ref{kincut}). The output L2 trigger rate was 
about 2.5\,kHz. The efficiency of the L2 trigger stage with respect to 
Level 1, averaged over the 2002 run, was measured to 
be $(83.7 \pm 2.1)\%$ for $\Xi^{0}$ $\beta$-decays, mainly
limited by wire inefficiencies in the drift chambers. 

The L2 trigger output rate was further reduced by about a factor 2 
at Level 3 (L3). The L3 trigger consisted of 
a farm of computers which used a specialized version of the off-line 
reconstruction code. It combined track measurements with 
clusters in the LKr calorimeter and used   
loose selection criteria. The inefficiency of the 
L3 trigger was measured to be less than
$0.1\%$. For normalization and efficiency determination 
purposes, the L3 trigger also received events from downscaled L1  
triggers as well as from NHOD pulses.

\section{Event selection and background rejection}

\subsection{$\Xi^0 \rightarrow \Sigma^+ e^- \overline{\nu}_e$}

The identification of the $\Xi^0 \to \Sigma^+ e^- \overline{\nu}_e$
channel was performed using the subsequent decay 
$\Sigma^+ \to p \pi^0$ with $\pi^0 \to \gamma \gamma$. 
The final state consisted of a proton and an electron leaving 
tracks in the spectrometer in addition to two photons being detected 
as clusters in the LKr calorimeter and one unobserved anti-neutrino. 
The decay $\Xi^0 \to \Sigma^+ \ell^- \overline{\nu}_e$ is the only source 
of $\Sigma^+$ particles in the neutral beam since the two-body decay 
$\Xi^0 \to \Sigma^+ \pi^{-}$ is kinematically forbidden. Thus, the 
signal events were 
identified by requiring an invariant $p\pi^0$ mass consistent with 
the nominal $\Sigma^+$ mass value.

The two tracks were required to be less than 2~ns apart in time, measured 
by the charged hodoscope or by the drift chambers when the hodoscope readout
was not able to reconstruct the track time. This occurred for about 2$\%$ of 
the events, 
mainly due to the presence of double-pulses from the scintillator 
photomultipliers. To suppress contamination from accidental activity in 
the detector, events with an additional track within a time window of 
20~ns with respect 
to the average time of the signal tracks were rejected.
  
\begin{figure}[htbp]
  \centering
  \begin{tabular}{c c}
    \includegraphics[width=7.9cm]{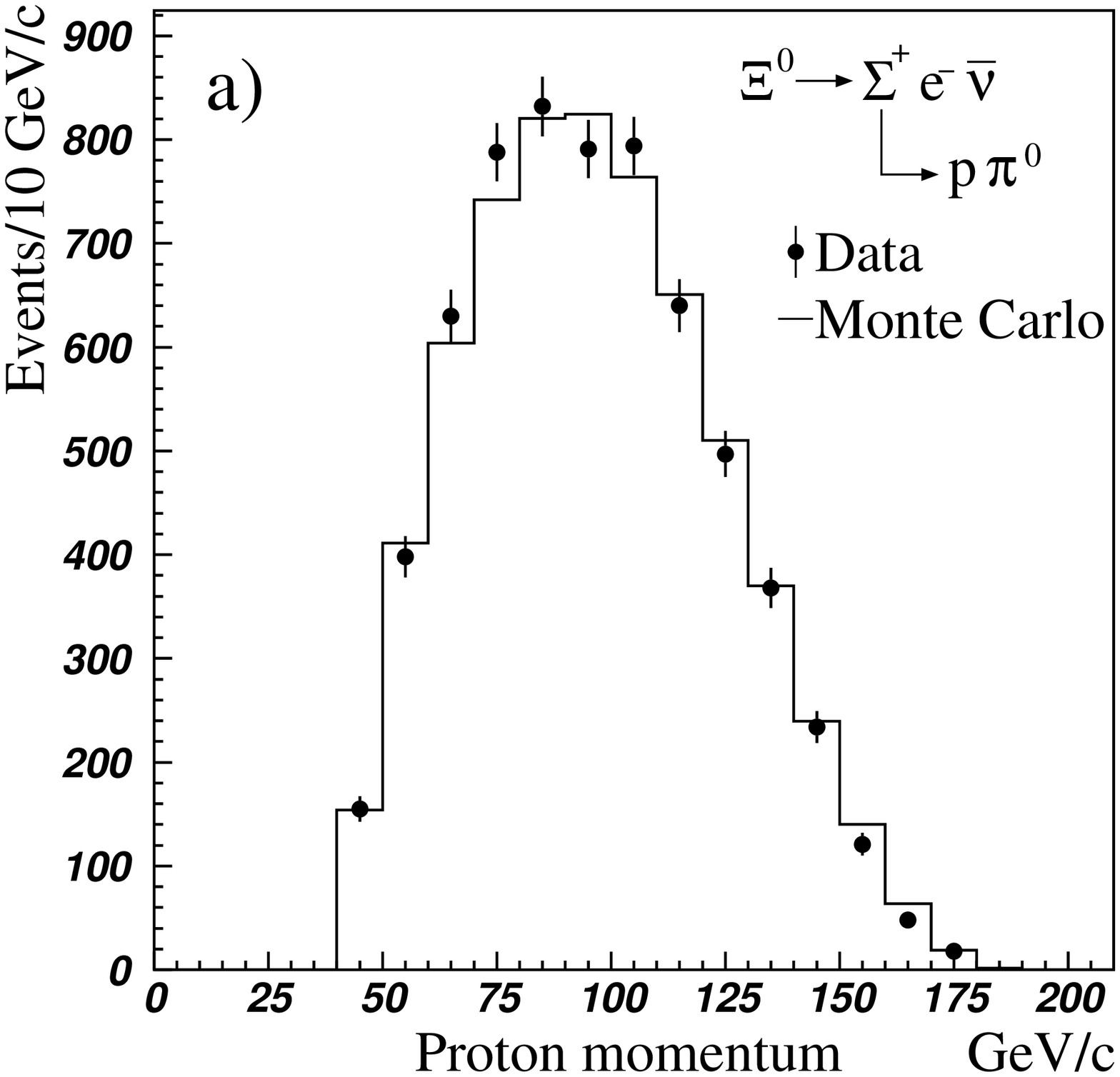}
    &
    \includegraphics[width=7.9cm]{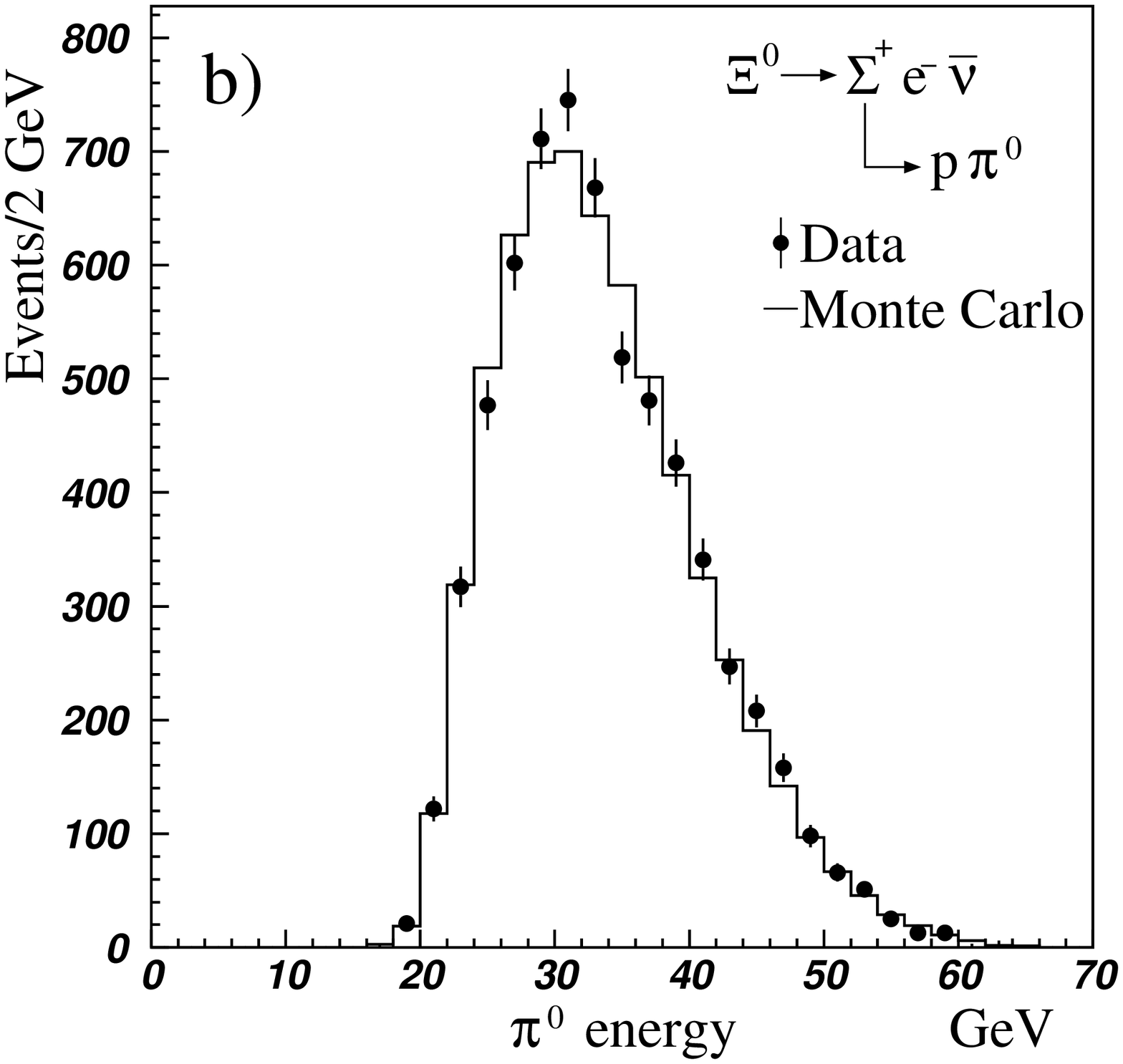}
  \end{tabular}
  \caption{Reconstructed proton momentum (a) and $\pi^{0}$ energy (b) 
distributions for $\Xi^{0} \rightarrow \Sigma^{+} e^{-} \overline{\nu}_{e}$
events with the subsequent $\Sigma^+ \rightarrow p\pi^0$ decay.} 
  \label{signal1}
\end{figure}

\begin{figure}[htbp]
  \centering
  \begin{tabular}{c c}
    \includegraphics[width=7.9cm]{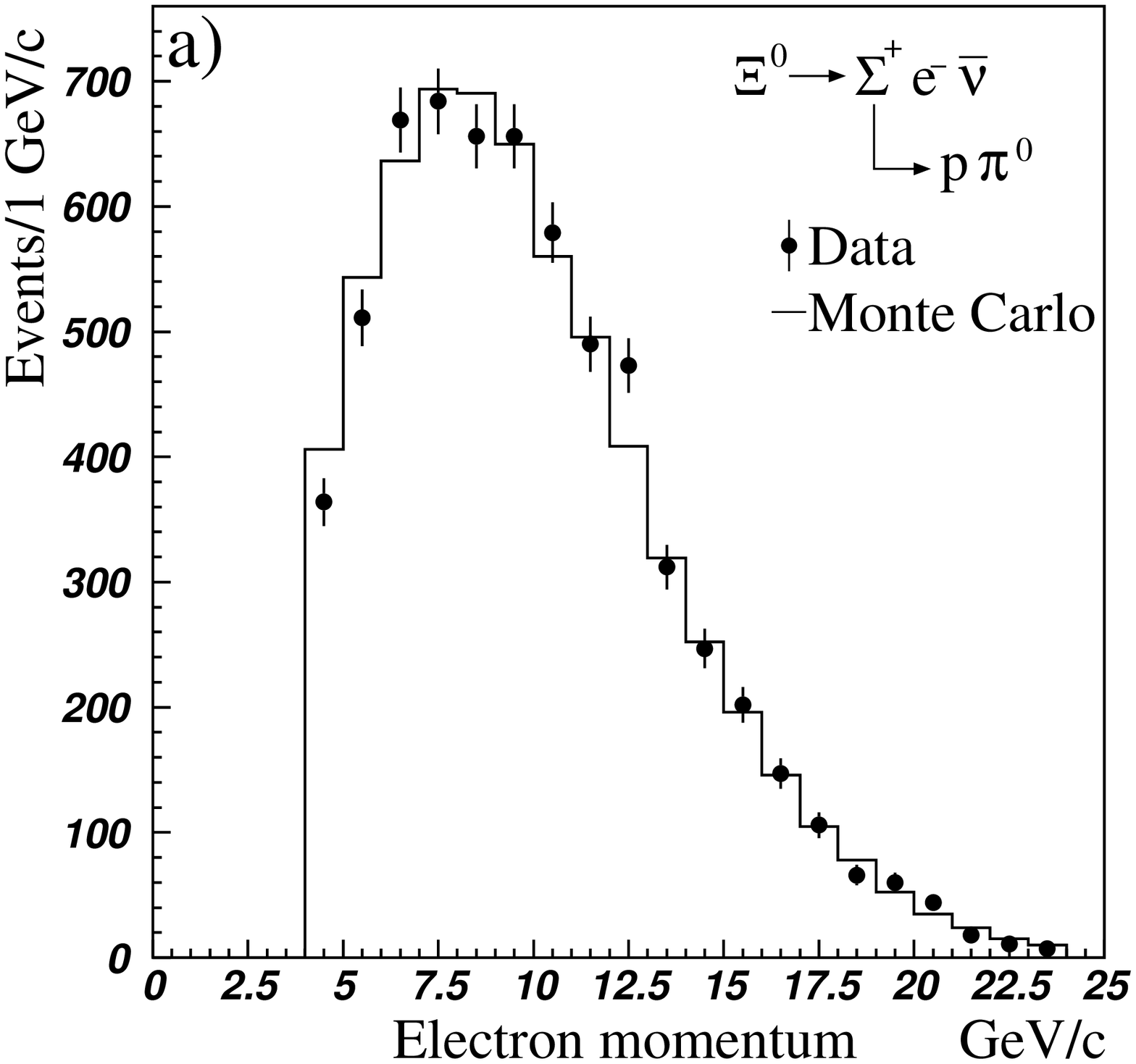}
    &
    \includegraphics[width=7.9cm]{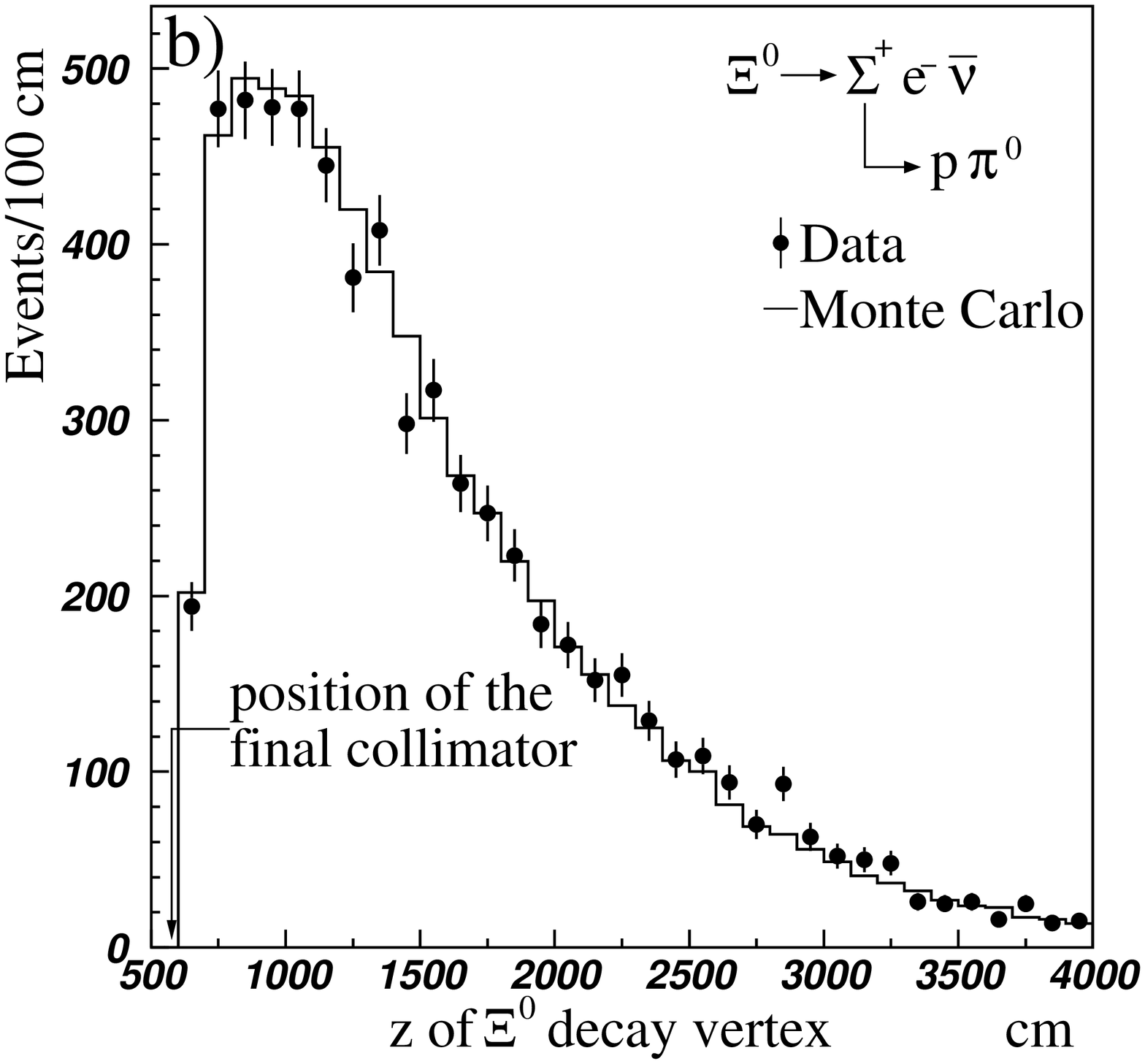}
  \end{tabular}
  \caption{Reconstructed electron momentum (a) and $z$-vertex coordinate (b) 
distributions for $\Xi^{0} \rightarrow \Sigma^{+} e^{-} \overline{\nu}_{e}$
events.}
  \label{signal2}
\end{figure}

The lower 
momentum thresholds for positive and negative tracks were set to 
40~GeV/c and 4~GeV/c, respectively (Fig.~\ref{signal1}(a) and 
Fig.~\ref{signal2}(a)). The momentum ratio between positive and 
negative tracks 
was required to be greater than 4.5 and the distance between the impact 
points of the two tracks in the first chamber was chosen to be greater 
than 12~cm in order to reduce biases from the corresponding cuts applied by L2. 
To ensure full efficiency in the track reconstruction, the radial 
distance to the beam axis of the reconstructed space points in the 
drift chambers had to lie between 12.5~cm and 110~cm. 

Electron identification was achieved by calculating the ratio $E/p$
of the cluster energy in the LKr calorimeter associated to the track with
the measured momentum 
in the spectrometer. Since electrons deposited their 
total energy in the electromagnetic calorimeter, their $E/p$ ratio was 
required to be between 0.85 and 1.15. For protons, the $E/p$ value was 
required to be less than 0.8. To avoid shower overlap, a minimum transverse 
distance of at least 15~cm was imposed between track impact points on 
the calorimeter surface.

The distance between the two tracks at the point of closest approach 
had to be less than 3~cm. 
In order to minimize the background 
coming from $\Xi^0 \to \Lambda \pi^0$ decays with $\Lambda \to p \pi^-$ and 
$\pi^{-}$ misidentified as electron, the difference between the 
nominal $\Lambda$ mass and the 
reconstructed invariant mass of the two tracks under the $p \pi^-$ hypothesis 
had to be greater than 14~MeV/c$^2$. Background from $K_S\to \pi^+ \pi^-$ 
decays with one accidental $\pi^0$ or two accidental photons was suppressed by 
rejecting events with an invariant $\pi^+ \pi^-$ mass within 30~MeV/c$^2$ 
of the nominal kaon mass and with a momentum ratio less than 6. 
The last three selection criteria were tighter than the 
ones used in the trigger to reduce 
biases from L2 trigger inefficiencies. 

The two-photon clusters forming a neutral pion candidate 
had to be within a time window of 2~ns and the energy of 
each cluster was required to be in the 3-100~GeV range. The 
reconstructed $\pi^{0}$ energy distribution is shown in 
Fig.~\ref{signal1}(b). 
Inner and outer regions of the LKr calorimeter were excluded by requiring
the radial distance to the beam axis of each cluster to be  
between 15~cm and 110~cm. Moreover, the center of each cluster was required
to be at a distance greater than 2~cm from any dead calorimeter cell.
To avoid biases in the energy measurement of the photons due to shower 
contamination induced by other particles, their associated clusters had to 
have a minimal distance from other clusters measured within a time window 
of 5~ns. 
This minimal separation was set to 10~cm for electron and photon candidates
and to 25~cm for hadronic showers associated with proton tracks. 
Photons originating from bremsstrahlung produced 
in the detector material before the magnet were rejected 
by measuring the separation at the LKr location between clusters and 
the impact point of the
extrapolated upstream segment of a track. 
%

The $\Sigma^+$ decay was reconstructed using a positive charged track 
in the spectro\-meter and two clusters in the electromagnetic 
calorimeter within a time window of 2~ns. The longitudinal position of 
the $\Sigma^+$ decay vertex 
was determined using the $\pi^0$ mass constraint to calculate
the distance of its decay point from the calorimeter:
\begin{eqnarray}
\Delta z_{\pi^0} = \frac{1}{m_{\pi^0}} \sqrt{E_1 E_2 r_{12}^2} 
\end{eqnarray}
where $E_1$, $E_2$ are the energies and 
$r_{12}$ the distance between  
the two clusters in the transverse plane of the calorimeter. The 
transverse position of the vertex
was then obtained by extrapolating back the proton track to the longitudinal 
position of the $\pi^0$ decay point. The momentum vector of the decaying
$\Sigma^+$ particle was calculated from the proton track parameters, the 
photon energies and assuming the emitted photons originate from the 
reconstructed vertex.      
 
The $\Xi^0$ decay vertex position was obtained by computing the 
closest distance of approach between the extrapolated $\Sigma^+$ 
line-of-flight and 
the electron track. This distance was required to be less than 4~cm. 
Furthermore, the deviation of the transverse $\Xi^0$ vertex position from 
the nominal line-of-flight defined by a straight line going from 
the center of the $K_{S}$ target to the 
center of the 
liquid krypton calorimeter was required to be less 
than 3~cm.   

The longitudinal position of the $\Xi^0$ vertex was required to be 
at least 6.5~m downstream of the $K_{S}$ target, i.e. 0.5~m after 
the end of the final collimator and at most 40~m from the target 
(see Fig.~\ref{signal2}(b)). Similarly, the $\Sigma^+$ vertex position was 
required to be at least 6.5~m downstream of the target but at most 50~m 
from the target. The latter value was chosen larger than the upper limit 
for the $\Xi^0$ vertex position to account for the lifetime 
of the $\Sigma^+$ particle. The longitudinal separation between the 
$\Xi^0$ and $\Sigma^+$ decay vertices was required to be between $-8$~m and 
$40$~m. The negative lower limit, tuned with Monte Carlo events, was chosen 
such as to take properly into account resolution effects.


The quantity $\vec{r}_{\mathrm{COG}}$
was defined as $\vec{r}_{\mathrm{COG}} = \sum_i \vec{r}_{i} E_i / \sum_i E_i$ 
where $E_i$ is the energy of the detected particle and $\vec{r}_{i}$   
the corresponding transverse position vector at the 
liquid krypton calorimeter position $z_{LKr}$. For a charged particle, 
the quantity $\vec{r}_{i}$ was obtained from the extrapolation to $z_{LKr}$ of 
the upstream segment of the associated track. The quantity 
$|\vec{r}_{\mathrm{COG}}|$ had
to be less than 15~cm. This requirement was found to produce negligible 
losses of signal events since the undetected neutrino in the 
$\Xi^0$ $\beta$-decay carries only a small 
fraction of the $\Xi^{0}$ energy.

Good candidates were kept if their $p\pi^0$ invariant mass was
found to be within 8\,MeV/c$^2$ of the nominal $\Sigma^+$ mass value, 
corresponding to a mass window of $\pm$4 standard deviations. Finally, the 
visible $\Xi^0$ energy was required to be in the 70 to 220~GeV
range. In the rare case that after all cuts were applied 
more than one candidate was found (more than one pair of photons 
associated to two tracks satisfying the event selection), 
the one with the smallest closest distance of approach between the $\Sigma^+$ 
line-of-flight and the electron track was chosen. 

\begin{figure}[htbp]
  \centering
    \includegraphics[width=10cm]{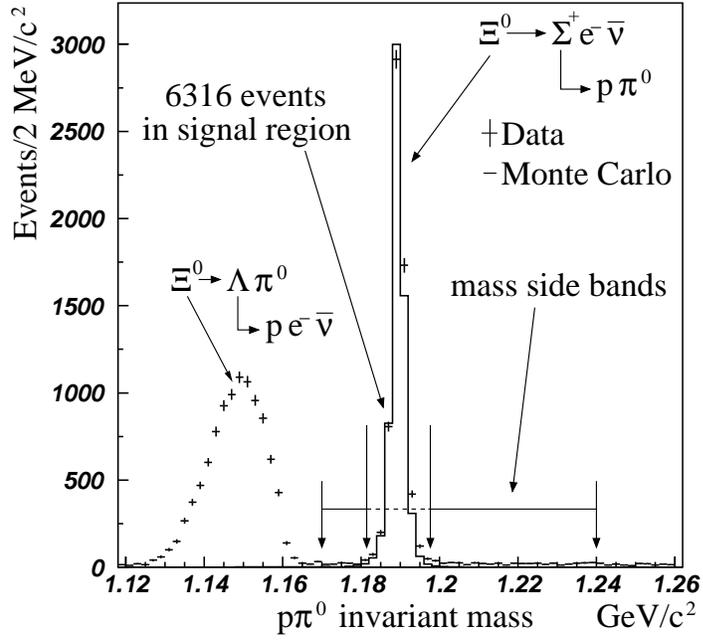}
  \caption{Reconstructed p$\pi^{0}$ invariant mass distribution 
for $\Xi^{0} \rightarrow \Sigma^{+} e^{-} \overline{\nu}_{e}$ candidates 
after all selection criteria were applied. The solid line shows the 
Monte Carlo prediction for the signal. The peak at the 
$\Sigma^{+}$ mass value shows clear evidence for the signal.
}
  \label{signal4}
\end{figure}

\begin{figure}[htbp]
  \centering
    \includegraphics[width=10cm]{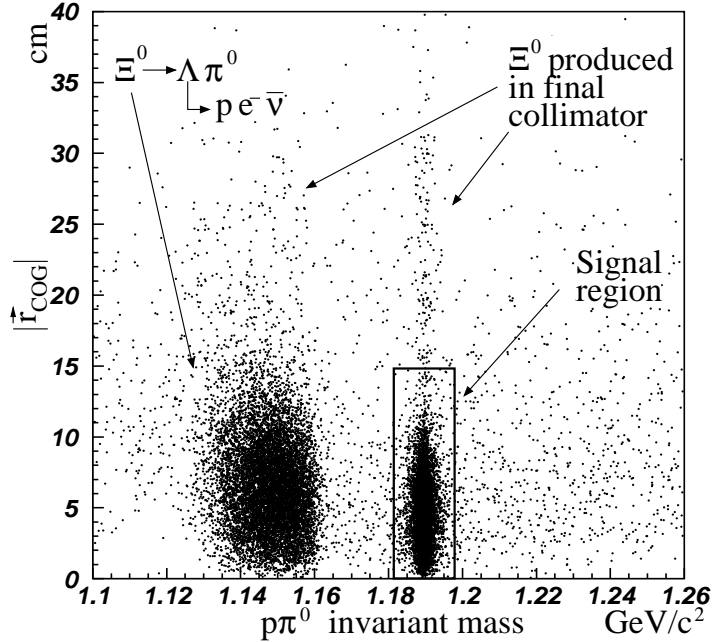}
  \caption{Scatter plot of $|\vec{r}_{\mathrm{COG}}|$ versus the p$\pi^{0}$ invariant mass for events
passing all other selection cuts. Events extending to high $|\vec{r}_{\mathrm{COG}}|$ values and centered around the $\Sigma^{+}$ mass are due to $\Xi^{0}$s produced in the final collimator.}
  \label{signal5}
\end{figure}

With the above selection criteria, 6316 
$\Xi^0 \rightarrow \Sigma^+ e^- \overline{\nu}_e$ candidates  
were observed in the signal region. The distribution of events in 
the $p\pi^0$ invariant mass variable 
is shown in Fig.~\ref{signal4} after all selection cuts were applied. 
Signal events peaking around the $\Sigma^+$ mass are clearly identified and
well separated from the abundant $\Xi^0\rightarrow \Lambda \pi^0$ decays 
(with $\Lambda \rightarrow p e^- \overline{\nu}_e$) located at low-mass
values. Monte-Carlo studies showed that contamination in the signal region 
from such events was negligible. Other background sources like
$K_L \rightarrow \pi^+\pi^-\pi^0$ decays or  
$\Xi^0\rightarrow \Lambda \pi^0$ followed by $\Lambda \rightarrow p \pi^-$ 
with mis-identified charged pions  
were also found not to contribute significantly. 
An amount of $(2.2 \pm 0.2)\%$ of  
background events in the signal region was estimated  
from the linear extrapolation of the distribution of events  
in the mass side-bands. By studying the time distribution of events in
side-bands regions, about $20\%$ only of this background was attributed
to residual accidental activity while most of the remaining 
contribution could be accounted for by re-scattering particles in 
the collimator, not rejected by the $|\vec{r}_{\mathrm{COG}}| < 15$\,cm selection cut. An 
additional source of unwanted events was associated with the 
production of $\Xi^0$s in the final collimator. Such events, although mostly 
present at large $|\vec{r}_{\mathrm{COG}}|$ values, exhibit a peak in the $p\pi^0$ 
invariant mass distribution, consistent with the $\Sigma^+$ mass 
(see Fig.~\ref{signal5}). 
Although these events are genuine $\Xi^0 \rightarrow \Sigma^{+} e^{-} 
\overline{\nu}_{e}$ decays, they were subtracted from the final sample 
in order to minimize uncertainties associated with their production yield 
and acceptance calculation. From inspection of the 
$|\vec{r}_{\mathrm{COG}}|$ distribution of events in the $\Sigma^+$ mass region, 
a contribution of $(1.2 \pm 0.7)\%$ of 
$\Xi^0$ $\beta$-decays originating
from the final collimator was estimated, yielding a total background 
contamination in the signal region of $(3.4 \pm 0.7)\%$.

\subsection{$\Xi^0 \rightarrow \Lambda \pi^0$}

To minimize systematic uncertainties in the branching ratio measurement,
the selection of the normalization events $\Xi^0 \to \Lambda \pi^0$   
with $\Lambda \to p \pi^-$ and $\pi^0 \to \gamma \gamma$ was performed
with analysis criteria as similar to the signal channel as possible. In 
particular, the same sensitive detector volume definition 
and time requirements for tracks and clusters were used. 

For the $\pi^-$ selection, the minimum momentum threshold was set to 5~GeV/c 
and no $E/p$ cut was applied.
Since the proton and the negatively charged pion originate from a
vertex ($\Lambda$ decay), the 
closest distance of approach between the two tracks was required to be less 
than 2.2~cm. The reconstructed invariant 
$p \pi^-$ mass was required to 
be within 4~MeV/c$^2$ of the nominal $\Lambda$ mass. The position
of the $\Lambda$ decay vertex was required to be at least 6.5~m downstream 
of the target but at most 50~m from the target.
As $\Xi^0 \rightarrow \Lambda \pi^0$ decays are fully reconstructed
in the detector, the upper value of 
the energy center-of-gravity was reduced to 7~cm.

\begin{figure}[htbp]
  \centering
  \begin{tabular}{c c}
    \includegraphics[width=7.9cm]{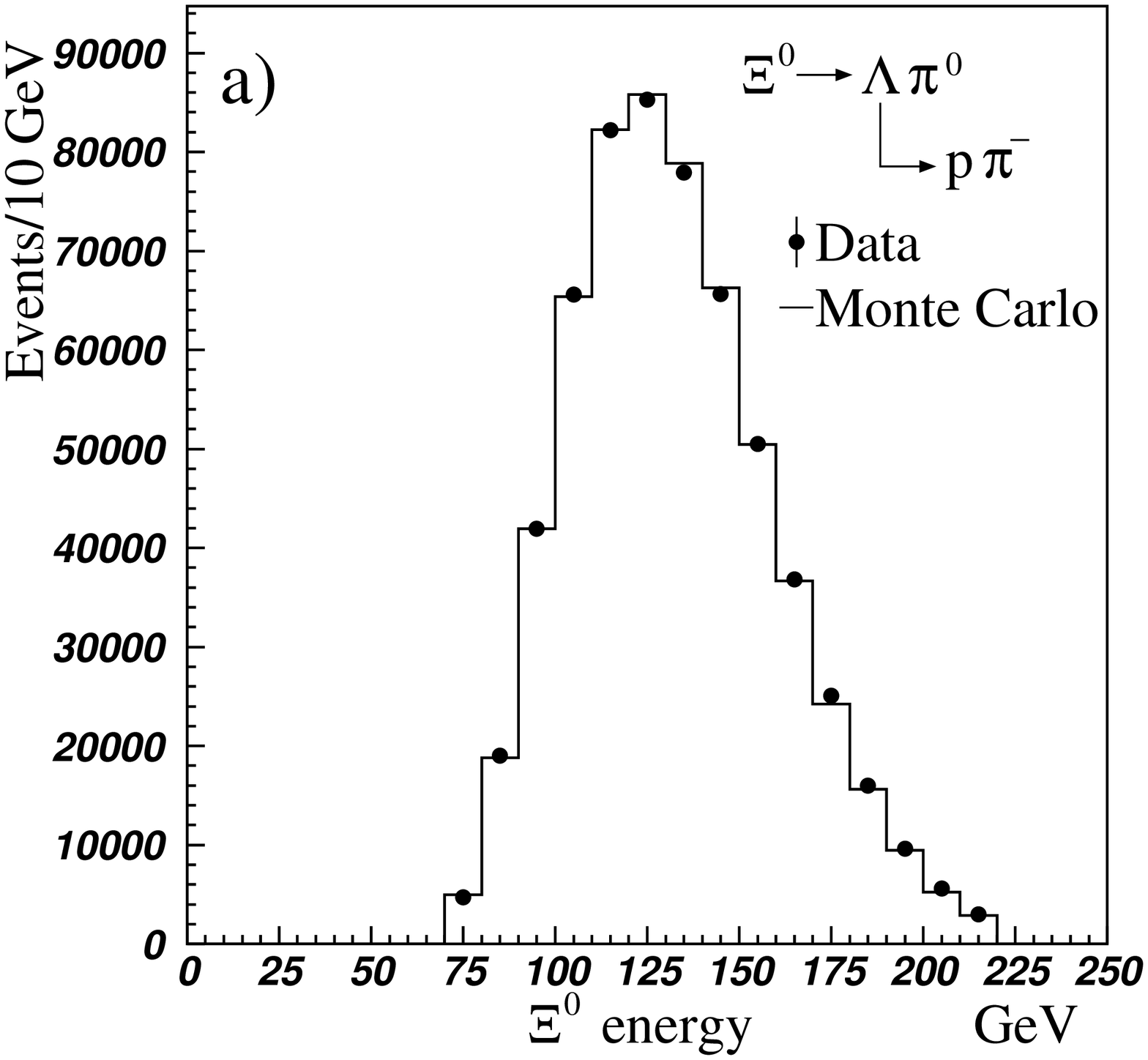}
    &
    \includegraphics[width=7.9cm]{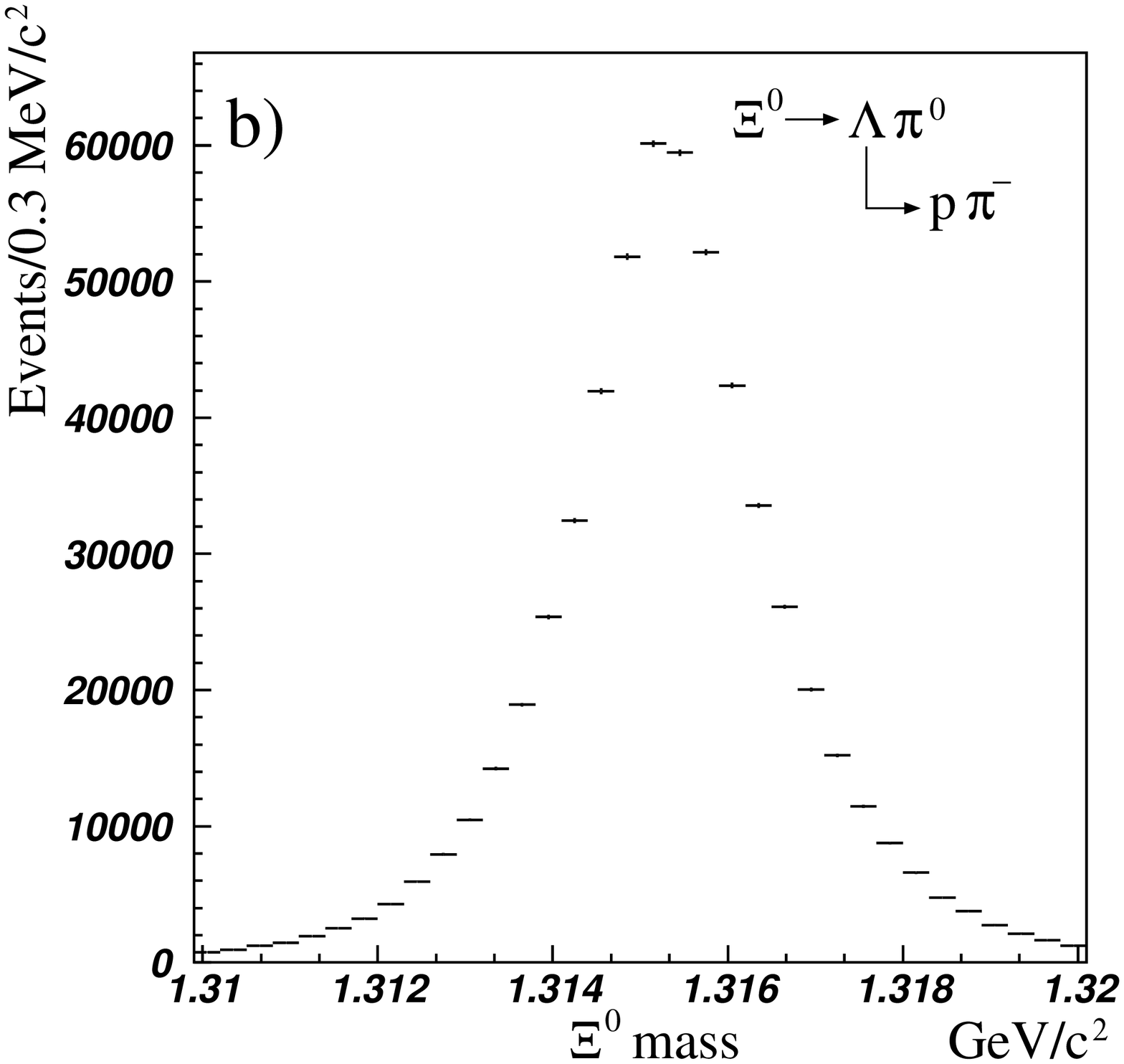}
  \end{tabular}
  \caption{(a) Reconstructed $\Xi^{0}$ energy and (b) $\Lambda \pi^0$ 
invariant mass distributions for $\Xi^{0}\rightarrow \Lambda \pi^{0}$ events.}
  \label{plotnorm}
\end{figure}

The longitudinal position of the 
$\Xi^0 \rightarrow \Lambda \pi^0$ decay point  
was defined by the $\pi^0$ vertex as in the case of the $\Xi^0$ $\beta$-decay 
by applying the same procedure for the vertex reconstruction. The fiducial
volume of the decay was contained longitudinally between 6.5~m and 40~m from 
the $K_S$ target and the $\Xi^0$ energy was required to be in the
70-220~GeV range (see Fig.~\ref{plotnorm}~(a)). Finally, the 
reconstructed $\Lambda \pi^0$ mass was required to be within 
the range 1.31 to 1.32~GeV/c$^2$. This mass window corresponds to about 
four standard deviations around the nominal $\Xi^0$ mass (see Fig.~\ref{plotnorm}(b)).

$588798$ candidates were observed in the signal region with a 
contamination of $(0.6 \pm 0.4) \%$ from $\Xi^0$s produced in the 
final collimator. After correcting for the average downscaling factor of 
33.79 applied to the L1 control trigger,  
the corresponding number of $\Xi^0 \to \Lambda \pi^0$ normalization events 
in the fiducial decay region was $1.990 \times 10^7$.

\section{Acceptance}

The acceptance for both signal and normalization 
decay channels was computed using a detailed Monte Carlo program
based on GEANT3~\cite{epsi,geant3}. Particle interactions in the detector material 
as well as the response functions of the different detector elements were 
taken into account in the simulation. 

\subsection{$\Xi^0 \rightarrow \Sigma^+ e^- \overline{\nu}_e$} 
\label{mcsignal}

The V-A transition matrix element for the decay 
$\Xi^{0} \rightarrow \Sigma^{+} e^{-} \overline{\nu}_{e}$
can be written as~\cite{cabswallwin}:

\begin{equation}
M = \frac{V_{\mathrm{us}}G_{F}}{\sqrt{2}}
[\bar{u}_{\Sigma^{+}}H_{\mu}u_{\Xi^{0}}][\bar{u}_{e}L^{\mu}u_{\nu}]+h.c.
\label{matrix}
\end{equation}   
where $V_{\mathrm{us}}$ is the appropriate
CKM matrix element for $|\Delta$S$|$=1 transitions, 
$G_{F}$ the Fermi coupling constant,
$\bar{u}_{\Xi^{0}}$, $\bar{u}_{e}$, $u_{\nu}$ and $u_{\Sigma^{+}}$
are Dirac spinors corresponding to the initial and final state particles.
$\bar{u}_{e}L^{\mu}u_{\nu}$ is the matrix element of the leptonic weak current
where $L^{\mu}$ has the well-established form
\begin{equation}
L^{\mu}=\gamma^{\mu}(1+\gamma_{5}) %
\end{equation}
and  $\bar{u}_{\Sigma^{+}}H_{\mu}u_{\Xi^{0}}$ is the contribution 
coming from the hadronic weak current.
The calculation of this term would require the treatment of
strong interaction effects. In practice, these are taken into 
account by introducing form factors in a parameterization of the most 
general form, compatible with Lorentz covariance:
\begin{eqnarray}
H_{\mu} = (O_{\mu}^{V}+O_{\mu}^{A})
\end{eqnarray}
with 
\begin{eqnarray}
  O_\mu^V &=& f_1(q^2)\gamma_\mu + 
  \frac{f_2(q^2)}{m_{\Xi^{0}}} \sigma_{\mu\nu}q^\nu +
  \frac{f_3(q^2)}{m_{\Xi^{0}}} q_\mu \\ 
  O_\mu^A &=& \left( g_1(q^2)\gamma_\mu + 
  \frac{g_2(q^2)}{m_{\Xi^{0}}} \sigma_{\mu\nu}q^\nu +
  \frac{g_3(q^2)}{m_{\Xi^{0}}} q_\mu \right) \gamma_5
\end{eqnarray}
In the expressions above, $f_{1}(q^2)$, $f_{2}(q^2)$ and $f_{3}(q^2)$ are the 
form-factors associated to the vector component of the hadronic weak current
while $g_{1}(q^2)$, $g_{2}(q^2)$ and $g_{3}(q^2)$ correspond to the 
axial vector part. The momentum transfer $q^2$, written in terms of  
the four-momenta of the involved particles, is:
\begin{eqnarray}
q^{\alpha}=(p_e+p_{\nu})^{\alpha}=(p_{\Xi^{0}}-p_{\Sigma^{+}})^{\alpha}
\end{eqnarray}
From the ingredients above, the 
$\Xi^0 \rightarrow \Sigma^+ e^- \overline{\nu}_e$ differential 
decay rate for a polarized initial hyperon beam could be calculated (for 
details, see~\cite{cabswallwin,garcia,linke}).

In the Monte Carlo simulation, we followed the prescription used 
in~\cite{cabswallwin} in which terms including the $f_{3}$ (scalar) and 
$g_{3}$ (pseudo-scalar) form-factors were neglected since they are 
suppressed in the transition amplitude by a factor
$m_{e}/m_{\Xi^{0}}$. In addition, the axial-tensor $g_{2}$ form-factor was
set to 0 as second class currents are forbidden
in the Standard Model. The values of the remaining non-vanishing 
form-factors $f_1$, $g_1$ and $f_2$ were obtained, with the 
assumption of SU(3) and CVC (conserved vector current) validity,
from the available data on neutron $\beta$-decay and the nucleon 
magnetic moments~\cite{PDG,gaillard}:  

\begin{eqnarray}
   f_{1}(0) &=& 1 \cr
   f_{2}(0)/f_{1}(0) = \frac{m_{\Xi^{0}}}{m_{n}} \frac{(\mu_{p}-\mu_{n})}{2}& = & 2.5966 \pm 0.0004 \cr
   g_{1}(0)/f_{1}(0) &=& 1.2695 \pm 0.0029,
\label{valoria0}
\end{eqnarray}
where $\mu_{p}$ and $\mu_{n}$ are the proton and neutron anomalous 
magnetic moments, respectively.
The above values for $g_{1}/f_{1}$ and $f_{2}/f_{1}$
are in good agreement with the ones directly measured 
from  $\Xi^{0}$ $\beta$-decays by the KTeV experiment~\cite{ktevff}:
\begin{eqnarray}
   g_{1}/f_{1} &=& 1.32^{+0.21}_{-0.17 \mathrm{stat}} \pm 0.05_{\mathrm{syst}} \cr
   f_{2}/f_{1} &=& 2.0 \pm 1.2_{\mathrm{stat}} \pm 0.5_{\mathrm{syst}} 
\label{valoriexp}
\end{eqnarray}

While $f_2$ was assumed constant, a dipole dependence as a function of 
the square of the 
momentum transfer was used for the $f_1$ and $g_1$
form-factors~\cite{cabswallwin,garcia}:

\begin{eqnarray}
f_{1}(q^{2})= f_{1}(0)(1+2\frac{q^{2}}{M_{V}^{2}})
\label{staf1}
\end{eqnarray}
\begin{eqnarray}
g_{1}(q^{2})= g_{1}(0)(1+2\frac{q^{2}}{M_{A}^{2}})
\label{staf2}
\end{eqnarray}
with $M_{V}=(0.97 \pm 0.04)$~GeV/c$^{2}$ and 
$M_{A}=(1.25 \pm 0.15)$~GeV/c$^{2}$~\cite{garcia, gaillard,cnops}.


The polarization of the $\Xi^{0}$ beam depends on both the hyperon
production angle and its momentum fraction with respect to the 
incoming proton beam. Since the $\Xi^{0}$ polarization was not measured 
in this experiment, an estimated value of $-10\%$ was used in the acceptance 
calculation. This amount is close to the preliminary measurement obtained 
by the KTeV experiment~\cite{ktevpol} for which the expected $\Xi^{0}$ 
polarization was comparable to the one in NA48/1. 

The subsequent $\Sigma^+ \to p \pi^0$ decay was simulated according to the 
well-known angular distribution for spin-1/2 hyperons decaying into
a spin-1/2 baryon and a pion~\cite{commins}:
\begin{eqnarray}
  \frac{d\Gamma}{d\Omega} =\frac{1}{4\pi} ( 1 + \alpha_{\Sigma^{+}} \: \vec{P}_{\Sigma^{+}}\cdot \hat{e})
\label{asymmetry}
\end{eqnarray}
where $\vec{P}_{\Sigma^{+}}$ is the polarization vector of 
the decaying $\Sigma^{+}$, 
$\hat{e}$ is the direction of the outgoing proton and
$\alpha_{\Sigma^{+}}=0.980^{+0.017}_{-0.015}$~\cite{PDG} 
is the corresponding asymmetry parameter of the decay.
 
Radiative corrections to the differential decay rate were included 
following the prescription of~\cite{garcia} in which model-independent 
contributions to first order in $\alpha$ from virtual and 
inner-bremsstrahlung graphs are taken into account in the 
transition amplitude.

The acceptance for the $\Xi^{0}$ $\beta$-decays  
in the fiducial decay region was calculated to be 
(2.492 $\pm$ 0.009)$\%$, where the quoted uncertainty originates 
from the statistics of the Monte Carlo sample. The inclusion of 
radiative corrections was found to increase the acceptance by $0.3\%$.

\subsection{$\Xi^{0} \rightarrow \Lambda \pi^{0}$}

The generation of the normalization events 
$\Xi^{0} \rightarrow \Lambda \pi^{0}$ 
with $\Lambda \rightarrow p \pi^{-}$ was performed 
using for each decay mode the form of the angular distribution given by
Eq.~\ref{asymmetry} with the appropriate values for the polarization vectors 
and asymmetry parameters. The $\Lambda$ polarization vector was obtained 
from the following relation:  
\begin{eqnarray}
  \vec{P}_{\Lambda}= \frac{( \alpha_{\Xi^0}+\vec{P}_{\Xi^0}\cdot \hat{e}) 
\cdot \hat{e}  + \beta_{\Xi^0}\cdot (\vec{P}_{\Xi^0}\times \hat{e})+
\gamma_{\Xi^0}\cdot \hat{e}\times (\vec{P}_{\Xi^0} \times \hat{e})}
{1+\alpha_{\Xi^0}\vec{P}_{\Xi^0} \cdot \hat{e}}
\end{eqnarray}
where $\hat{e}$ is the direction of the outgoing $\Lambda$, $\vec{P}_{\Xi^0}$ 
the polarization vector of the initial hyperon and $\alpha_{\Xi^0}$= -0.411,
$\beta_{\Xi^0}$=0.327 and $\gamma_{\Xi^0}$=0.85 are the asymmetry parameters 
used in the simulation for the 
$\Xi^{0} \rightarrow \Lambda \pi^{0}$~\cite{PDG}. For the 
non-leptonic $\Lambda \rightarrow p \pi^{-}$ channel,
the value $\alpha_{\Lambda} = 0.642$~\cite{PDG} was used.

The acceptance for the normalization 
$\Xi^{0}\rightarrow \Lambda \pi^{0}$ events in the fiducial decay region
was found to be (1.377 $\pm$ 0.004)$\%$, assuming a polarization of -10$\%$
for the initial $\Xi^{0}$. The quoted uncertainty on the acceptance is
again purely statistical.

\section{$\Xi^0 \rightarrow \Sigma^+ e^- \overline{\nu}_e$ branching ratio}
The determination of the $\Xi^0 \rightarrow \Sigma^+ e^- \overline{\nu}_e$ 
branching ratio was obtained from the background subtracted 
number of good events for signal and normalization, the
corresponding acceptance values,  
the L2 trigger efficiency measured with respect to the L1 one and    
the normalization branching ratios~\cite{PDG}. These quantities 
are summarized in Table~\ref{brcalc} and yield: 
\begin{eqnarray}
\mathrm{BR}(\Xi^0 \rightarrow \Sigma^+ e^- \overline{\nu}_e) =(2.51 \pm 0.03_{\mathrm{stat}}\pm 0.09_{\mathrm{syst}})\times10^{-4}
\end{eqnarray}
where the statistical uncertainty originates from the event statistics and 
the systematic one is the sum in quadrature of the 
various contributions presented in Table~\ref{systematics}.
\begin{table}[htbp]
\begin{center}
\caption{Parameters used for the 
BR($\Xi^0 \rightarrow \Sigma^+ e^- \overline{\nu}_e$) measurement.\label{brcalc}}
\begin{tabular}{| l | c | c|}
\hline
& $\Xi^0 \rightarrow \Sigma^+ e^- \overline{\nu}_e$ &
$\Xi^0 \rightarrow \Lambda \pi^{0}$ \\
\hline
 Event statistics & $6316$ & $588798$ \\
 Downscaling factor   & 1  & $33.79$ \\
 Background & $(3.4 \pm 0.7)\%$ & $(0.6 \pm 0.4)\%$\\
 Acceptance &  $(2.492\pm0.009)\%$ & $(1.377\pm 0.004)\%$ \\
\hline
 L2/L1 trigger efficiency &  $(83.7 \pm 2.1)\%$ & \\
\hline
BR$(\Sigma^+ \rightarrow p \pi^0)$ & $(51.57\pm0.30)\%$& \\
BR$(\Xi^0 \rightarrow \Lambda \pi^0)$ & &$(99.523\pm0.013)\%$ \\
BR$(\Lambda \rightarrow p \pi^-)$ & &$(63.9\pm0.5)\%$ \\
\hline
\end{tabular}
\end{center}
\end{table}
This result is in good agreement with existing 
measurements~\cite{ktev2}. 
 
The largest contribution to the total systematic uncertainty 
comes from the L2 trigger efficiency whose determination was limited in
precision by the statistics available in the control samples.
The sensitivity of the branching ratio measurement to the form factors
was studied by varying $f_2/f_1$ and $g_1/f_1$ within the limits
provided by the uncertainties on $M_V$, $M_A$ masses and on   
the $f_2(0)/f_1(0)$, $g_1(0)/f_1(0)$ parameters (Eq.~\ref{valoriexp}), and
was found to be mainly dominated by the precision on $g_1$.
  
Other systematic uncertainties due to acceptance and selection criteria
were estimated to be  $1.0\%$
by varying the geometrical and kinematical
cuts applied in the event selection. In particular, the sensitivity
of the branching measurement to the inner radius cut in the first drift
chamber was investigated. Since in both signal and normalization channels the 
outgoing protons carry a large part of the primary $\Xi^{0}$ energy,
an important fraction of them travels along the detector near 
the beam pipe, in a region where the spectrometer efficiency and acceptance
may change rapidly. The corresponding systematic uncertainty obtained 
from the variation of the measured branching 
ratio as a function of the inner radius cut in the first drift chamber
was estimated not to exceed 0.6$\%$. 
  

\begin{table}[htbp]
\begin{center}
\caption{Sources of systematic uncertainties.\label{systematics}}
\begin{tabular}{| l | r |}
\hline
Source & uncertainty \\
\hline
 
 Background                               & $\pm0.8\%$ \\
 MC statistics                            & $\pm0.5\%$ \\
 L2 trigger efficiency                    & $\pm2.2\%$ \\
 Form factors                             & $\pm1.6\%$ \\
 Geometrical and kinematical cuts         & $\pm1.0\%$ \\
 $\Xi^0$ polarization                     & $\pm1.0\%$ \\
 $\Xi^0$ lifetime                         & $\pm0.2\%$ \\
 Normalization                            & $\pm1.0\%$ \\
\hline
 Total                                    & $\pm3.4\%$ \\ 
\hline
\end{tabular}
\end{center}
\end{table}

Finally, an uncertainty of $\pm 5\%$ in the polarization 
of the initial $\Xi^{0}$ was assumed, resulting in a contribution of $1.0\%$
to the systematic uncertainty on the 
$\Xi^0 \rightarrow \Sigma^+ e^- \overline{\nu}_e$ branching ratio.  

\section{$\overline{\Xi^0}\to\overline{\Sigma^+}e^+ \nu_e$ decays}

\begin{figure}[htbp]
  \centering
  \begin{tabular}{c c}
    \includegraphics[width=7.9cm]{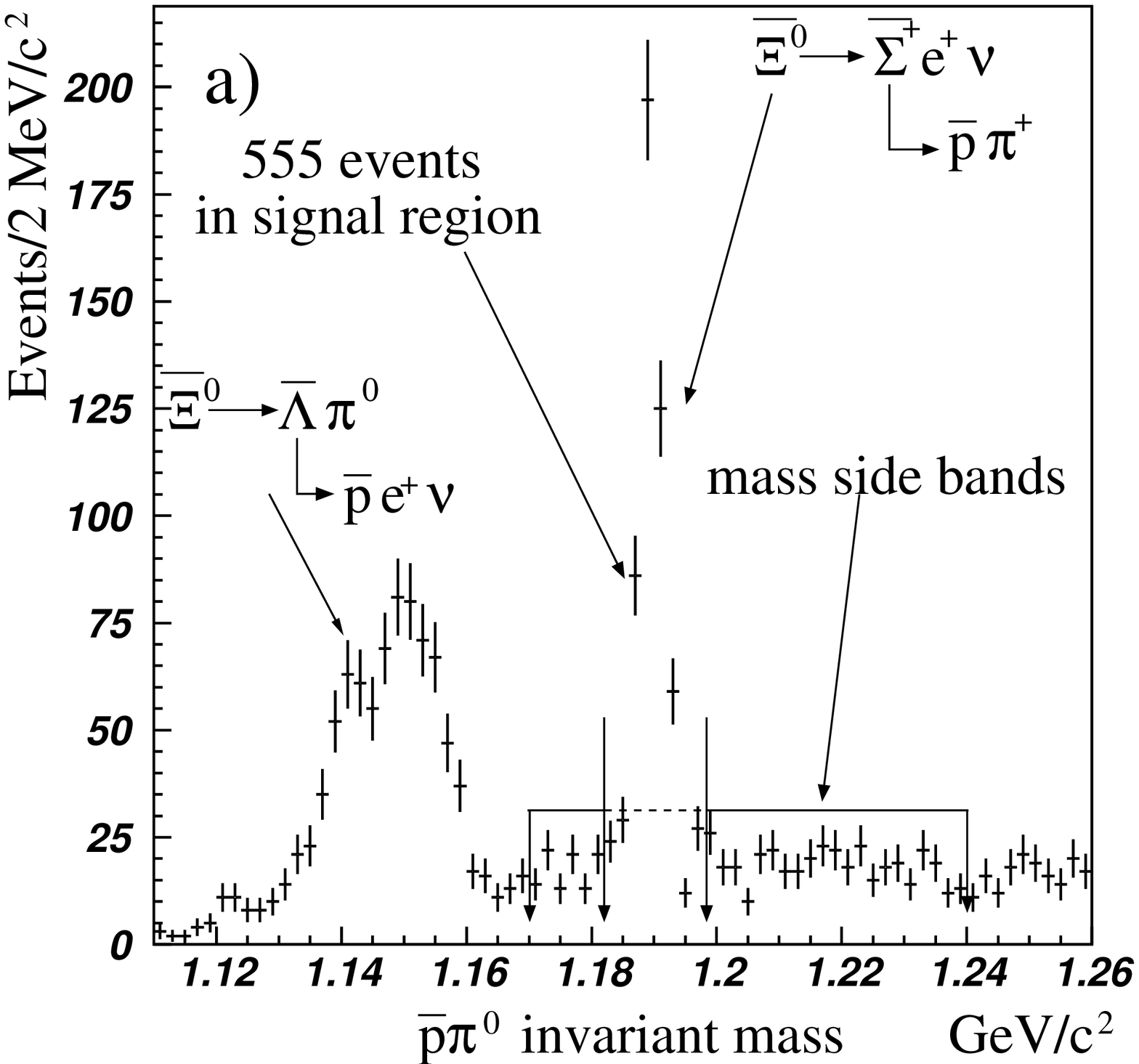}
    &
    \includegraphics[width=7.9cm]{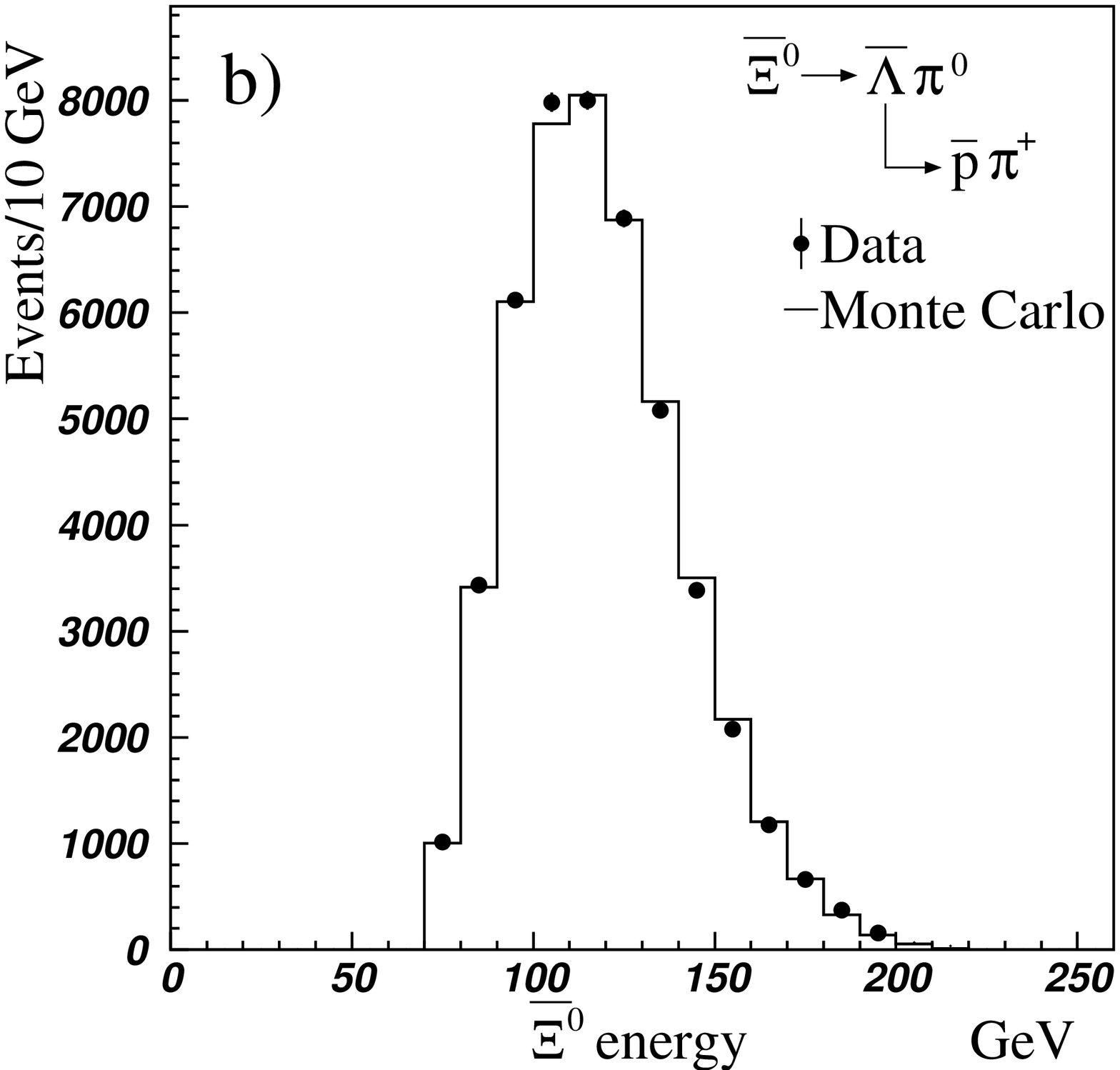}
  \end{tabular}
  \caption{(a) $\overline{\mathrm{p}}\pi^{0}$ 
invariant mass distribution of events. The peak corresponding to the 
$\overline{\Sigma^{+}}$ mass shows 
clear evidence for $\overline{\Xi^{0}}$ $\beta$-decays. 
(b) $\overline{\Xi^{0}}$ energy distribution from 
$\overline{\Xi^{0}}\rightarrow \overline{\Lambda} \pi^{0}$ events.}
  \label{fig:anti_signal}
\end{figure}

Since the trigger system did not distinguish between particle charges 
with respect to the event hypotheses, the recorded data sample also 
contained decays of anti-hyperons, allowing the first measurement of the 
$\overline{\Xi^0}\to\overline{\Sigma^+}e^+\nu_e$ branching ratio to be 
performed. Events originating from 
$\overline{\Xi^0}\to\overline{\Lambda}\pi^0$ decays were used as
the normalization channel. In order to minimize systematic differences between 
the branching ratio determinations for $\Xi^0$ 
and $\overline{\Xi^0}$ $\beta$-decays, the same selection criteria for both
modes were applied with the exception of the required charge 
inversion for tracks. Fig.~\ref{fig:anti_signal}(a) shows the 
$\overline{p}\pi^0$ invariant mass distribution of events after all other cuts 
were applied. A sample of 555 $\overline{\Xi^0}\to\overline{\Sigma^+}e^+\nu_e$ 
candidates was found in the signal region with a background contamination of 
$136 \pm 8$ events, measured from the extrapolation of the 
flat distribution of 
events in the side-band regions around the $\overline{\Sigma^+}$ mass and taking into account possible contributions from $\overline{\Xi^0}$s
produced in the final collimator.
For the normalization channel $\overline{\Xi^0}\to\overline{\Lambda}\pi^0$,
$47351$ events with negligible background were identified using 
data samples obtained from control triggers. After taking into account the 
downscaling factors applied to the recorded data, the corresponding  
number of $\overline{\Xi^0}\to\overline{\Lambda}\pi^0$ events in the 
fiducial decay region was found to be $1.601 \times 10^6$.  

The acceptance calculation was performed using Monte Carlo samples 
that were generated with the same matrix element as 
for the study of $\Xi^0$ $\beta$-decays
and with a production energy spectrum adjusted to fit the spectrum
of observed $\overline{\Xi^0}\to\overline{\Lambda}\pi^0$ events
(see Fig.~\ref{fig:anti_signal}(b)). The $\overline{\Xi^0}$ production 
polarization was set to zero, as expected for anti-hyperons,
and the signs of the decay parameters for both signal and normalization
channels were changed according to the theory. 
The acceptance was found to be $1.80\%$ for the 
$\overline{\Xi^0}\to\overline{\Sigma^+}e^+\nu_e$ decay mode 
and $1.19\%$ for the normalization channel, with negligible statistical 
uncertainties. 

The L2 trigger is the main component of the trigger 
which affects the semileptonic $\overline{\Xi^0}$ branching ratio measurement. 
Due to the limited number of reconstructed 
$\overline{\Xi^0}$ events from control triggers, the L2 trigger efficiency 
was assumed to be the same as for $\Xi^0$ $\beta$-decays. However, 
an additional systematic uncertainty of $2.0\%$ was added in quadrature 
in order to account for possible effects due to the  
different $\overline{\Xi^0}$ polarization value and production spectrum.  

The branching ratio for 
$\overline{\Xi^0}\to\overline{\Sigma^+}e^+\nu_e$ decays was measured to be:
\begin{equation}
\mathrm{BR}(\overline{\Xi^0}\to\overline{\Sigma^+}e^+ \nu_e) = (2.55 \pm 0.14_{\mathrm{stat}}\pm{0.10}_{\mathrm{syst}}) \times 10^{-4}
\end{equation}
in very good agreement with the value obtained above 
for $\Xi^0$ $\beta$-decays. The 
relative systematic uncertainty of $3.9\%$ is dominated by the trigger 
efficiency determination. Contributions to the systematic uncertainty from
form factors, geometrical cuts and acceptance, rescattering effects in the 
final collimator as well as normalization were 
obtained from the study of the semileptonic $\Xi^0$ decay.

\section{Determination of $|V_{\mathrm{us}}|$ and $g_1/f_1$}

The $|V_{\mathrm{us}}|$ parameter can 
be extracted from the measured $\Xi^0$ semileptonic decay rates using the
following relation~\cite{garcia}:
    
\begin{equation}
\begin{split}
\Gamma=\frac{BR_{\Xi^{0}\rightarrow \Sigma e \nu}}{\tau_{\Xi^{0}}} 
= &G_{F}^2 |V_{\mathrm{us}}|^{2} \frac{\Delta m^{5}}{60\pi^{3}}(1+\delta^{\mathrm{MD}}_{\mathrm{rad}})(1+\delta^{\mathrm{MI}}_{\mathrm{rad}}) \\
&\times \biggl\{(1-\frac{3}{2} \beta)(|f_{1}^{2}|+3|g_{1}^{2}|) 
+\frac{6}{7}\beta^2(|f_{1}^{2}|+2|g_{1}^{2}|+
Re(f_{1}f_{2}^{*})+
\frac{2}{3}|f_{2}^{2}|) \\
&+\delta_{q^{2}}(f_{1},g_{1})\biggr\} 
\end{split}
\label{vusextra}
\end{equation}
where $\tau_{\Xi^{0}}=(2.90 \pm 0.09)\times 10^{-10}$~s is the 
$\Xi^{0}$ lifetime,
$\Delta m=m_{\Xi^{0}}-m_{\Sigma^{+}}=0.12546 \pm 0.00021$~GeV/$\mathrm{c}^{2}$
and $\beta=\frac{\Delta m}{m_{\Xi^{0}}}= 0.09542 \pm 0.00011$~\cite{PDG},
$\delta^{\mathrm{MD}}_{\mathrm{rad}}= 0.0211$ and $\delta^{\mathrm{MI}}_{\mathrm{rad}}= 0.0226$ are, respectively, 
model-dependent  
and model-independent radiative corrections and
$\delta_{q^{2}}(f_{1},g_{1})=0.119$ takes into account 
the contribution from the transfer 
momentum dependence of the form-factors $f_1$ and $g_1$~\cite{garcia}.
Eq.~\ref{vusextra} was computed neglecting terms of $O(\beta^{3})$. 

Using the combined result $BR_{\Xi^{0}\rightarrow \Sigma e \nu}
=  (2.51 \pm 0.09)\times 10^{-4}$ of the measured $\Xi^0$ and 
$\overline{\Xi^0}$
branching ratios together with the current experimental 
determination of
$g_1/f_1$ and $f_2/f_1$~\cite{ktevff} and neglecting SU(3) breaking 
corrections to $f_1$, the value for 
$|V_{\mathrm{us}}|$ was found to be 
\begin{equation}
|V_{\mathrm{us}}| = 0.209^{+0.023}_{-0.028}~,
\label{vusvalue}
\end{equation}
consistent with the present value obtained from kaon semileptonic
decays~\cite{PDG}. The uncertainty on $|V_{\mathrm{us}}|$ is dominated by 
the experimental precision on $g_{1}/f_{1}$, and the corresponding 
contribution 
due to the branching ratio measurement itself is now comparable 
to the error on the $\Xi^{0}$ lifetime. 

Conversely, the $g_{1}/f_{1}$ ratio could be extracted from Eq.~\ref{vusextra}
using the current $V_{\mathrm{us}}$ value 
obtained from kaon 
decays~\cite{PDG}: 
\begin{equation}
g_{1}/f_{1}=1.20\pm 0.04_{\mathrm{br}} \pm 0.03_{\mathrm{ext}}
\end{equation}
where the uncertainty coming from the present branching ratio
measurement (br) takes into account the weak dependence
of the acceptance on $g_{1}/f_{1}$ itself.
The external error (ext) includes the contributions from $V_{\mathrm{us}}$, 
$\Xi^{0}$ lifetime
and $f_2/f_1$ uncertainties.
Our measurement is in agreement with exact SU(3) symmetry and 
favours theoretical approaches in which SU(3) breaking effects 
do not modify significantly the $g_{1}/f_{1}$ ratio.
  
\section{Conclusion}
Using the data collected in 2002 with the NA48 detector at CERN, we 
obtained the first determination of the  
$\overline{\Xi^{0}}\rightarrow \overline{\Sigma^{+}} e^{+} \nu_{e}$ branching 
ratio and performed a measurement of the $\Xi^{0}\rightarrow \Sigma^{+} e^{-} \overline{\nu}_{e}$ branching ratio with a 
precision significantly better than the existing
published values. Our results provide, in addition, a new
determination of the ratio $g_{1}/f_{1}$ or, alternatively, of
the $|V_{\mathrm{us}}|$ parameter.   

%
%
%
\section*{Acknowledgments}
It is a pleasure to thank the technical staff of 
the participating laboratories,
universities and affiliated computing centers for their efforts in the 
construction of the NA48 apparatus, in the operation of the experiment, and in 
the processing of the data.

\newpage



\end{document}